\documentclass{article}

\usepackage{PRIMEarxiv}

\usepackage[utf8]{inputenc} % allow utf-8 input
\usepackage[T1]{fontenc}    % use 8-bit T1 fonts
\usepackage{hyperref}       % hyperlinks
\usepackage{url}            % simple URL typesetting
\usepackage{booktabs}       % professional-quality tables
\usepackage{amsfonts}       % blackboard math symbols
\usepackage{nicefrac}       % compact symbols for 1/2, etc.
\usepackage{microtype}      % microtypography
\usepackage{lipsum}
\usepackage{fancyhdr}       % header
\usepackage{graphicx}       % graphics
\graphicspath{{media/}}     % organize your images and other figures under media/ folder
\usepackage{xcolor}         % For colored comments
\usepackage{authblk}        % for better formatting of author block
\title{Interactive and Urgent HPC: \\ Challenges and Opportunities}

\author{Albert Reuther}
\affil{MIT LL}
\author{Nick Brown}
\affil{EPCC}
\author{William Arndt}
\affil{LBL}
\author{Johannes Blaschke}
\affil{LBL}
\author{Christian Boehme}
\affil{GWDG}
\author{Antony Chazapis}
\affil{FORTH}
\author{Bjoern Enders}
\affil{LBL}
\author{Robert Henschel}
\affil{IU}
\author{Julian Kunkel}
\affil{GWDG, Universität Göttingen}
\author{Maxime Martinasso}
\affil{CSCS}

\begin{document}
%\sptitle{Article Type: Description  (see below for more detail)}
\maketitle

%\markboth{THEME/FEATURE/DEPARTMENT}{THEME/FEATURE/DEPARTMENT}

\begin{abstract}

As a broader set of applications from simulations to data analysis and machine learning require more parallel computational capability, the demand for interactive and urgent high performance computing (HPC) continues to increase. This paper overviews the progress made so far and elucidates the challenges and opportunities for greater integration of interactive and urgent HPC policies, techniques, and technologies into HPC ecosystems. 

\end{abstract}

% keywords can be removed

\section{Introduction}
\label{sec:introduction}

Computing is fundamentally interactive; it's how almost all of the world interacts with their computers. 
However, highly parallel research and scientific computing, commonly termed high performance computing (HPC)/supercomputing (usually used interchangeably), in which tens, hundreds or thousands of high-end servers are used in parallel to solve massive simulation, data analysis and machine learning training tasks, are just starting a transition to greater interactivity. 
Since supercomputers started emerging about 50 years ago, they have always been seen as very high value systems, similar to expensive experimental equipment like telescopes, that require proposals for allocations, planning, and meticulous scheduling to execute jobs against management-granted compute time allocations. Furthermore, the types of simulations that were, and still are, executed on these supercomputers have required extensive analysis and planning to determine the parameters, initial conditions, and meshes with which to execute each parallel job. 
This type of execution is commonly referred to as batch scheduling. 
In the past two decades however, it has become apparent that a significant subset of HPC jobs are more effective for the user when they are run without waiting in a queue. These jobs are typically known as interactive HPC and urgent HPC, and their common denominator is being sensitive regarding their start or/and completion times. 

Interactive HPC involves users \emph{being in the loop} during job execution where a human is monitoring a job, steering the experiment, or visualizing results to make immediate decisions about the results to influence the current or subsequent interactive jobs. 
Interactivity is often the first step in scaling to larger models or datasets, as well as being part of an agile development and testing cycle. 
It is not only computational workloads, but also data analytics and machine learning workflows that frequently require interactive exploration of large data sets. 

Likewise, urgent HPC involves immediate data or actions that will fail if the job is not run with immediacy/urgency. Urgent HPC is similar to deadline scheduling, where an HPC job has to deliver results by a given time. However, urgent HPC jobs require the job execution to start immediately. Further, urgent HPC jobs can be planned or unplanned, depending on whether the event to which the job is couple can be planned or not. 
A rapidly growing area of HPC usage is in urgent supercomputing for tackling emergency scenarios, for instance the extensive use of supercomputing during the global pandemic and recent bouts of extreme climate events. These have demonstrated the need to make urgent, accurate decisions for complex problems, and combining interactive computational modelling with the near real time detection of unfolding disasters results in a powerful tool that can aid emergency responders making life-critical decisions for disaster response.  In a similar vein, supercomputers can be used in the operating room guiding the surgery process, provided urgency requirements are met\cite{Zhou2020}.  Ultimately the objective here is to exploit HPC to deliver significant societal benefits by saving lives and reducing economic loss. 

While numerous attempts around the interactive and urgent use of HPC resources have been undertaken, and this field has enjoyed some progress, yet the HPC community has been slow to adopt interactivity in a widespread and consistent fashion. This is one of the reasons why we strongly believe that such a series of workshops is important, because, not only do we need to grow the interactive HPC community in order to demonstrate the benefit to the domain scientists, reduce tool complexities, and challenge the traditional usage models of resource access, but furthermore we have already started to see impact in these areas based upon our activities to date. 
And much progress has been made in bringing interactive and urgent HPC capabilities online at a number of supercomputing centers across the world. 

In Section~\ref{sec:currentstate}, we take inventory of the current state of the practice in enabling and implementing interactive and urgent HPC capabilities across the world. 
This inventory includes the current state of developing policies, examining system metrics, gathering data, enabling schedulers, developing tools and environments, providing user support, and highlighting user case studies. 
The inventory then elucidates gaps and opportunities for further research that will enable future interactive and urgent HPC capabilities, which are discussed in Section~\ref{sec:opportunities}. Finally, a conclusion will draw this paper to a close. 
\section{The State of the Practice}
\label{sec:currentstate}

In this section, we capture the current state of the practice in regards to interactive and urgent HPC capabilities. 
Over the past several years, numerous birds-of-a-feather meetings and workshops have been hosted at IEEE/ACM Supercomputing and ISC-High Performance conferences, and from which we will synopsize the current state of the practice.
The current state is divided into the following topics: organizational and system policy; scheduling techniques; system infrastructure and tools; data management; user support; and user case studies. 

\subsection{Organizational and System Policy}

When a supercomputing center considers introducing or expanding the number of compute nodes that are available for executing interactive and urgent sessions, they are usually faced with several organizational and management policy concerns. 
These policy concerns are rooted in the traditional view that purchasing and maintaining an HPC system is very expensive and that jobs on the HPC system must be scheduled such that the system being fully utilized (over 95\% is common) to justify its purchase and recurring operations costs. To keep an HPC system fully utilized, jobs are scheduled in ``batch mode'', where users submit their jobs into a queue of pending jobs, which wait until resources are available to execute each one. Wait times can range from minutes to hours to days and weeks depending on many factors. Consequently, batch scheduling does not accommodate interactive and urgent jobs where idle nodes must be available to quickly launch such jobs, as this leads to lower utilization percentages.  
Some centers have introduced new metrics based on return on investment (ROI), replacing the utilization metric that is more focused on user productivity~\cite{reuther2006high}. The DARPA HPCS program in the early 2000s, from which the aforementioned paper came, and a BoF session at SC-13 on cost-benefit analysis of HPC tried addressing this policy concern, but limited progress has been made since then.

\subsection{Scheduling Urgent and Interactive Jobs}

How jobs are scheduled on HPC systems plays a large role in their suitability for interactive and urgent workloads. All HPC systems have at least one shared login node upon which users can execute some interactive work. If a system is large enough, then there is usually a debug queue which provides interactive debugging, but there are usually few resources allocated to that queue and sessions are usually short, 30 minutes to two hours maximum, depending on the center. None of these adequately accommodate full interactive and urgent jobs of any scale and duration.  
Some centers use reservations to provide resources that can accommodate interactive and urgent sessions where the debug queues are not suitable. Other centers go further by identifying certain jobs that can be preempted, and these share a queue with interactive and urgent jobs~\cite{byun2020best}, typically with owners of such preemptable jobs given preferential treatment (e.g. prioritised scheduling or reduced cost). 

Along the same lines, much of the published work on scheduling interactive and urgent jobs on HPC clusters has focused on the currently-prominent schedulers, Slurm and Kubernetes, with early efforts also leveraging HTCondor, IBM Platform LSF, GridEngine, PBS, and Torque schedulers. For the most part, Slurm has become the defacto standard HPC scheduler, while Kubernetes has become the standard cloud/container scheduler for clusters. 

Research has evaluated various feature sets for launching interactive jobs with regard to the types of jobs that these schedulers support, for instance high throughput computing and interactive jobs versus large scale, optimally mapped jobs~\cite{reuther2018scalable}. 
It has been found that consistently there are tensions between how quickly large jobs are scheduled versus smaller/interactive jobs. Further tension exists between multi-node synchronously parallel jobs and block synchronous (high throughput) jobs. A good balance between all of these is difficult or perhaps impossible to find. 

This tension has driven the development of strategies and implementations that make interactive and real-time software appear like batch jobs, but operate interactively. For instance, BatchSpawner enables Jupyter notebook servers to run in a batch job, and KubeSpawner allows them to run as a pod under Kubernetes on demand, the later arguably being a more natural approach which has seen greater adoption. Pilot jobs \cite{turilli2018comprehensive} that load tasks from an external database and process these create the illusion of queue-less real-time analysis as a service, but handling this within the queue system itself would be a better approach. 

Another approach has been implemented in Kubernetes, where a new scheduler is created and sends jobs to optimized sub-schedulers for different types of jobs depending on a variety of characteristics such as job size. Different queues and priorities are exploited to effectively schedule these jobs, and a similar feature is being developed for the Flux scheduler, which aims to succeed Slurm. 

Urgent jobs with latency requirements in the one digit second range, such as LLM inference in chatbots, cannot usually be accommodated by single job scheduling. Typically, the scheduling task required is demand based scaling of a permanently available service. This type of scheduling is provided by Kubernetes autoscaling, and to integrate this service scheduling with classical HPC job scheduling on the same system, different approaches have been developed. For example, Kubernetes has been extended by the batch scheduler Volcano and there is ongoing work on provisioning Slurm clusters under Kubernetes using the MPI Operator or other similar services.

\subsection{Technology for an Interactive and Urgent Service}

Effective scheduling and execution policies are fundamental for facilitating interactive and urgent computing, yet they are insufficient in isolation. Equally crucial are the environments and tools which have been specifically designed to support interactive and urgent workflows. This requirement is addressed through interactive and programmatic environments tailored for workflow construction. Users engaged in interactive and urgent computing tasks typically require environments that enhance their productivity, including code development, debugging, and data analysis. Such environments can be categorized into integrated application environments, which facilitates application management; portals that provide web-based access to HPC resources tailored for domain-specific tasks; and APIs that enable the development, customization and automation of complex computational workflows.

Initial efforts to establish an interactive and urgent user environment involved the creation of custom solutions by HPC providers. A notable instance of such an environment is the configuration of a MATLAB framework on a computing cluster, utilizing MatlabMPI and pMatlab libraries for parallel processing. These MATLAB processes were initiated via the cluster's job scheduler~\cite{reuther2007technical}. Concurrently, analogous systems such as Star-P, gridMathematica, and Techila were developed, each aiming to provide users with the tools necessary for interactive computation and data processing in an HPC context.

The progression of specialized HPC tools has given rise to more versatile development environments, such as customizable JupyterLab instances and feature-rich versions of Visual Studio Code. These environments, while powerful, require specific expertise and resources to configure for particular applications. In response, several HPC centers have independently developed portals to simplify access to such tools. Early examples include the portals created by the Department of Defense High Performance Computing Modernization Program (DoD HPCMP), MIT Lincoln Laboratory, and the San Diego Supercomputer Center (SDSC), which typically utilize proxy-based mechanisms to provide desktop users with browser-based access to integrated tools such as JupyterLab.
Building on these initiatives, researchers at the Ohio Supercomputing Center and the University of Virginia introduced Open OnDemand~\cite{OpenOnDemand}, an open-source framework. This framework furnishes HPC centers with the underlying structure and portal capabilities required to offer a diverse array of interactive and urgent HPC services via a web interface. Open OnDemand serves as a pivotal development in democratizing access to HPC resources, streamlining the user experience.

Alternative approaches have facilitated the integration of research desktops which are environments provisioned directly on HPC systems, providing seamless access to HPC services. The research desktop paradigm allows users of varying expertise to leverage computational power and storage capacity with the familiar ease of a traditional desktop interface.
These research desktop servers function similarly to the login nodes of an HPC system where they are situated in close proximity to HPC clusters, have cluster file systems mounted, and act as submission hosts for batch processing. Unlike login nodes, which typically offer only SSH access, research desktop nodes provide remote desktop access using solutions such as ThinLinc, NoMachine, X2Go, or FastX. Additionally, they are configured to allow the execution of sustained graphical applications, such as MATLAB, RStudio, or Visual Studio Code, thus accommodating a range of computational tasks within a user-friendly environment.

In parallel, some organizations have pioneered the development of web-based programmatic interfaces to HPC resources, empowering the scientific community to construct bespoke portals and workflow engines. These HPC centers offer RESTful API access to their services, principally enabling data transfer and job submission functionalities. Utilizing these APIs, along with their language-specific bindings, researchers can seamlessly integrate interactive and urgent HPC workflows, customizing them to optimize productivity.
An example is the Materials Cloud~\cite{MaterialClouds} platform, which facilitates interactive studies of material properties through a dedicated web portal. It harnesses the AiiDA workflow engine
written in Python and the Python bindings of FirecREST
a web-facing APIs. Such APIs simplify support requirements for HPC centers as they can provide a single interface to cater to diverse scientific requirements.
Additionally, various other APIs like HEAppE, the Superfacility API, and HPCSerA have been developed. Some APIs also provide serverless computing endpoints, which allow for service-based calls, further extending the functionality and flexibility of HPC resources for user-driven computational tasks.

Another ongoing theme is the convergence of HPC and Cloud computing. From the HPC perspective, there are a breadth of tools and services already available for the Cloud in HPC such as HPK (High Performance Kubernetes) which allows users to deploy Kubernetes within an HPC environment. Using a custom kubelet, this translates the container's lifecycle to Slurm and Singularity commands. Moreover, concepts are introduced into the HPC world to provide Cloud flexibility on top of HPC systems. An example of this is the model of versatile software-defined cluster~\cite{vcluster}. By leveraging network segregation and containerization, this model offers a more flexible and adaptable service delivery to HPC users, aligning with Cloud computing paradigms while maintaining the core strengths of vertically integrated software stack of HPC systems.

\subsection{Data Management}
\label{sec:curr_data_management}

In many cases (\emph{cf.} Section~\ref{sec:introduction}) the need for urgent and interactive resources arises from the analysis of data (often from external sources). Therefore effective data management can become a lynchpin of any urgent and interactive workflow. HPC facilities are aware of this, and have started to develop infrastructure to better manage high volumes and rates of data. Here we list those areas which we have identified as most impactful to urgent and interactive HPC workflows.

Today, most HPC facilities deliver data storage and management through deployment of large shared file systems along with a mechanism that maps their user landscape to Unix users and file groups. 
Additionally, object storage interfaces such as S3 are provided as cold storage or for data sharing.
The engineering trade-offs that are made to enable large-scale shared storage can lead to unpredictable performance, which, in turn, can cause workflow stalls due to 100x slower performance then expected. 
While many teams' workflow jobs will be impacted by slow I/O, users with interactive or urgent workflows are impacted  particularly severely as important deadlines would be jeopardized \cite{blaschke2021realtime}. 
As these users require reliable access to compute resources, the same naturally applies for all I/O associated with the interactive/urgent workflow. 
This concern on the quality of service for data I/O can readily be expanded to all networks that are touched in the path of the data flow. For many urgent use cases, where data is generated externally but analyzed locally, the data need to be shipped to the HPC facility in time for the compute job to run. Waiting for data to arrive is akin to waiting in the scheduler queue for nodes.
The unreliable nature of shared file systems and the asynchronicity associated with file transfer to and from the HPC storage infrastructure has forced some teams that prioritize fast feedback look into memory-to-memory streaming solutions where shared filesystems are avoided until the end of the pipeline. However, these solutions are novel from an HPC point of view, and might require exceptions to firewall rules and thus face resistance from a security and policy perspective at HPC facilities. Furthermore, as some HPC systems are designed with no direct path between compute nodes and external networks, such designs will not only introduce additional latency, but also additional potential points of failure.

Another issue is the coordination of data transfer with computation. While some workflow management tools have begun to develop features and best practises, the current landscape is dominated by each application implementing its own solution with little guidance from HPC centers. These solutions normally take the form of regularly checking the state of the file system, or polling an API as to the state of a given transfer. This almost always happens independently of the HPC center's batch scheduler. This lack of coordination with the HPC center's batch scheduler results in additional delays in starting data analysis, and leaves tracking data provenance entirely to the workflow tool.

Finally, interactive and urgent workflows face challenges with HPC data management around more of a social or trust nature across scientific domains. For instance, data at experimental facilities are often generated by machine rather than a person, and may be attributed to a group of people that don't have a corresponding file group at the HPC facility or that don't even have user accounts. Moreover, that team could be of a transient nature and dissolve after the data has been analyzed and published. Intense competition in the sciences also forces them to embargo the data until it's ready for publication. HPC facilities today are not well equipped to serve this user group as projects and group mappings may be made through a separate process where the Principal Investigator (PI) or group for the HPC accounts differs from the PI or group that "owns" the data. The strategies to map their social groups to HPC groups or users vary. One group might opt to have everything done under a single "machine collaborative" account to which there exists a separate access mapping for a few select admin accounts. Another strategy is to go all-in and have every external user apply for an HPC account. Both approaches are not ideal, as the former raises questions about acceptable use and traceability of user actions while the latter greatly inflates the number of users on an HPC system and puts the management burden onto HPC staff.

\subsection{System User Support}

Interactive and urgent HPC capabilities attract both traditional and unconventional users including biology/chemistry users, social sciences, etc. Unconventional users often are not used to scaling up their simulations, experiments, etc., nor are they familiar with the usual software tools and environments of HPC. Similarly, traditional HPC users may not be as familiar with the interactive and urgent HPC environments, techniques, and tools. The question is how to get these users up to speed on pertinent tools quickly to give them success and motivation to explore further. For the unconventional users, in particular, becoming proficient and confident in using HPC systems can seem daunting.  

As with any HPC center, expert and patient computational scientists and engineers who work as research facilitators are essential to pointing users to further online content and examples. The research facilitators should be helping users solve their issues, and not solving them for the users. (Teach users how to fish, rather than giving them another fish each time they ask.) 
For common questions and issues, boilerpate emails and online content (knowledge bases) can reduce the load on research facilitators, while getting users acclimated to searching and accessing online content~\cite{mullen2018lessons}.  
Further, consortium content such as HPC Carpentry~\footnote{\url{https://www.hpc-carpentry.org/}} provide shared resources for teaching basic HPC skills. 
The HPC Certification Forum\footnote{\url{https://www.hpc-certification.org/}} is an effort to identify and organize competencies to clearly define the semantics of the competences for practitioners, trainers and learners - however, urgent computing is not yet covered.
Some HPC centers are using online courses to teach HPC basics including helping users configure their environment (including laptop/desktop) to take advantage of the HPC capabilities. 
And some centers are requiring completion of online courses to get full default resource allocations. Such online courses can also be used to gain HPC certifications, which can be important for users in their career progression.

\section{Research Challenges and Opportunities}
\label{sec:opportunities}

\subsection{Organizational and System Policy}

Attitudes within the HPC community rooted in a resource scarcity view result in a focus on maximizing system utilization as the metric of success for HPC systems.
But the obvious way to enable interactive and urgent workflows on an HPC system is at odds with this metric: Provision supplementary hardware capacity dedicated to time-sensitive jobs and accept that resources may idle to ensure that such jobs start promptly.
Here it may be useful to reframe the question and investigate what are the truest indicators of scientific productivity for an HPC system.
While acknowledging that HPC resources are finite, we believe interactive and urgent workflows in HPC offer a different dimension of scientific value that the state-of-the-art approach in maximizing system utilization fails to capture.

To a scientist whose real-time data analysis requires HPC, a system that cannot deliver an answer within time constraints imposed by sample availability, instrument time allocation, or funding simply has no scientific value.
From this perspective, HPC centers are focused on optimizing the wrong thing.
Therefore, \textbf{we advocate for the reporting of HPC center metrics that characterize key requirements of time-sensitive workflows} like prompt queue wait time, network latency, and I/O subsystem performance variance alongside traditional utilization metrics.
These measurements can be used to set goals through technology and policy choices that balance time-sensitive computing needs and traditional batch workloads.

Also, as individual HPC systems are subject to maintenance and potential faults there always must be an alternative center to pickup urgent computational jobs. 
This stresses the requirement for portability and organizational measures to ensure a smooth fall back execution in case the initially targeted center is unavailable.

\subsection{Effective Scheduling}

It is generally acceptable that batch scheduling, as implemented today, are insufficient for interactive and urgent HPC workloads. Batch scheduling provides unbounded determinism, where the requested resources are guaranteed to be provided but there are no constraints around when they will become available. Interactive scheduling, by contrast, must provide a minimum set of resources within a specific time frame and be able to flexibly scale with the changing resource requirements of interactive work. 
Furthermore, in urgent scheduling we need to reliably provide the requested resources with known latency and also to provide redundancy in case of failures. %The common denominator of effective scheduling for interactive and urgent jobs is therefore low latency allocation of resources.

It makes sense to pursue technical solutions that enable batch, interactive, and urgent job scheduling to co-exist on the same system. Slurm's support for preemption enables the system to free up resources on demand for interactive and urgent workflows, although in practice stopping jobs and cleaning up queues may incur long delays. Moreover, preemption requires well-defined policies on which jobs will be interrupted that preserves fairness and ensures completion of long-running tasks.

Even though schedulers such as Slurm are constantly being developed to accommodate various degrees of flexibility, scheduling decisions need to account for more than job size and queue priority. These metrics should be able to change dynamically in time, and job malleability allows reservations to grow (or shrink) to facilitate prompt job commencement while maximising resource utilization. However, this depends on support from both the runtime and the application, and introduces overheads mapping jobs to the new set of resources.
The batch scheduler could also offer the ability to allocate any resources available using a best effort approach, or support some kind of resource negotiation protocol, which factors in the expected duration of the job.
As many urgent tasks are not unanticipated, this is an opportunity to co-schedule resources across facilities. Co-scheduling can leverage negotiation protocols, policy, and facility APIs, which are in development now.
Response time metrics could also be fed back to the system so that it can ensure a specified level of interactivity. Such feedback loops have been used extensively in the Cloud to help the system balance resource allocations to required performance.

A method to handle conflicting scheduling policies is to simultaneously apply different algorithms on the same resources. Because high resource utilization is less important than low latency for interactive and urgent HPC, service schedulers, such as Kubernetes, are arguably better suited to this than traditional batch schedulers such as Slurm. 
On a multi-purpose HPC system it is desirable for these two paradigms to co-exist and an example of this is where some HPC systems  reserve supplementary resources for time-sensitive workflows, effectively dynamically partitioning the system. Flux embraces the idea of multiple scheduling contexts, dynamically adjusting the cluster in multi-level nested partitions running different schedulers including Kubernetes.  In the Cloud, Kubernetes supports many schedulers to operate simultaneously on the same resources and it is up to the jobs to decide which one to use. It is yet to be verified if multi-scheduler solutions can handle interactive and urgent jobs alongside batch processes, or some unified solution can deliver the required balance between all job types. In any case, finding ways for batch schedulers and containerized service orchestration to run alongside one another, or enable one on top of the other, will become increasingly important in coming years.

Following these assumptions we identify the following gaps requiring additional research:

\begin{description}
    \item[Dynamic out-scaling:] While service schedulers are already providing dynamic out-scaling mechanisms, they still need to be integrated with the interactive or urgent workloads. For data analytics and machine learning there already exist suitable integrations (via frameworks as Dask or Pytorch Elastic) but efficient integration of dynamic MPI scaling for parallel jobs with service schedulers in a production system is a topic of ongoing research.
    \item[Integrating batch and low latency scheduling:] To allow for both batch and low latency scheduling, approaches are to extend service schedulers with batch scheduling (e.g., Volcano for Kubernetes) or to make batch scheduling faster and better scalable (Flux scheduler), or to run a service scheduler side-by-side with a batch scheduler. 
    Significant efforts are required to provide sustainable and productive solutions.
    \item[Freeing resources:] Job preemption is the tool of choice to make (additional) resources immediately available for interactive and urgent jobs. For this we need to improve the share of HPC jobs which can be reliably suspended. Furthermore, schedulers need to select suitable candidates for preemption based on policy, priority, practicality, and efficiency loss. 
    \item[Redundancy:] For urgent jobs deadline assertions may be in place, which require measures to recover from resource failures. 
    This may necessitate redundant scheduling, i.e., running a copy of the workload.
\end{description}

\subsection{Technologies and Tools}

Open OnDemand, JupyterHub, proxy services, remote desktop servers for GUIs, and web dashboards today give users access to HPC, bypassing the Linux command line that many researchers find unfamiliar, intimidating, or just not the right fit for scientific workflows.
Extrapolating that this trend of bringing familiar user interface paradigms to HPC will continue, we expect that future researchers will expect to interact with HPC through tablets or phones, and through new modalities like natural language.
This opens up new opportunities to make HPC more usable and accessible, but will require HPC developers to center user experience and human-computer interaction --- a new area of research for HPC.

Distributed, interactive workflows that incorporate HPC resources into a real-time analysis loop with remote instrumentation expand the scope of the programming endeavor, highlighting new and exciting problems to solve.
HPC software developers are familiar with performance analysis and optimization tools for applications, but what does the equivalent for a distributed analysis workflow look like?
At a minimum, each segment of a workflow including data acquisition, transfer, storage, compute, analysis, and archiving needs to advertise its status accurately, in sufficient detail, and in a timely fashion.
HPC components need to advertise subsystem status using standard machine-readable formats with greater granularity than just ``system up'' or ``system down.''
Standardizing and exposing these metadata streams is both a programming challenge and a sociological one.
But solving that problem will lead to new kinds of distributed, interactive workflow applications that bring HPC to bear on big-data science problems.

Finally, interfaces using protocols new to HPC environments like HTTP and websockets pose new challenges for networking and security.
Organizations that previously only had to provision Linux accounts for users and monitor ssh-based access to login nodes will have to contend with federated identity, analyzing web traffic, and configuring external network routes to compute nodes.
Today's researchers grew up with real-time collaborative productivity tools like Google Docs, but the closest analogy in a typical HPC context is account sharing --- entirely incompatible with HPC security models.
How can the needs and expectations of users be reconciled with HPC center infrastructure, networking, and security requirements?
Institutions that provide HPC and their vendors will need to work together to identify standards-based technologies from the broader market and open-source community that have addressed these challenges in other contexts.
Articulating the need for new capabilities and adapting solutions to HPC are other areas where contributions can be made.

\subsection{Data Management}

The data management state of the practise outlined in section~\ref{sec:curr_data_management} shows that -- while HPC infrastructure is capable of handling large volumes of raw data -- more research, development, and improvements to center policy are needed in order to allow urgent and interactive HPC users to effectively and reliably deploy their workflows on HPC. This work has the highest chance of succeeding, if it is a combined effort among the whole HPC community: HPC center staff, researchers, users, and vendors. We have identified the following areas of research:

\begin{description}
    \item[Reliable High-Speed Data delivery:] Data needs to arrive and be available to the computational resources (e.g. visible on the compute nodes) quickly and reliably. To this end we propose the implementation of quality of service (QoS) for networking and storage. This has two aspects: prioritization of traffic and I/O for time-critical applications; and minimal performance guarantees in order to allow users to plan their tasks appropriately.
    \item[Effective Data Transport through Streaming:] While many workflows are served well by viewing data storage as POSIX file systems, this approach introduces additional latency. We propose that the HPC community include data streaming services (direct to compute node memory) as part of their standard data offering. By providing a standardized service, this can be integrated with QoS, and automatic backup and fault tolerance services.
    \item[Center-Wide Data Management:] We must avoid that users of urgent tasks also need to manage all aspects of data transfer and storage. We therefore propose that the community develop a uniform API which can: i) provide information on data transfers; ii) enable users to register event hooks (e.g. calling a function after a file has been transferred); iii) track data provenance (e.g. which files where read during a job); and iv) request/modify network and storage QoS. Additional tools, such as real-time performance and data flow monitoring, I/O tuning, and helping users organize paths can be built upon such an API.
    \item[Collaborative Access Models:] many urgent and interactive are collaborative in nature. The HPC community needs to develop an access scheme that covers the uses cases described in section~\ref{sec:curr_data_management}. These range from collaborative spaces (where a team can work on shared data sets), to fine-grained control over who has access to which data set. This scheme needs to be controllable by trusted individuals (who are not necessarily HPC sysadmins -- e.g. experiment operators), and fast (within minutes -- e.g. in order to quickly bring in an external expert). We note that modern data centers have tiered storage, and related data is often spread over these. This access model must therefore cover all storage tiers in the data center.
\end{description}

In addition to the technical research areas listed above, we note that there is a social challenge of combining practices in computing science with those in experimental science. Often this leads to friction, especially in the areas of policy and security. Since the best designed tool will have limited usefulness if its operation violates data center policy (e.g. data streaming to compute nodes is not useful if external connections to compute nodes are prohibited by security policy), a great deal of work needs to be done in order to align urgent and interactive data management requirements and data center policy. 

\subsection{User Education and Training}

Urgent and interactive HPC is one pathway by which users from non-computing disciplines are first introduced to HPC. Therefore many interactive and urgent HPC users start out being rather inexperienced. Hence, there are many opportunities to encourage their enthusiasm for HPC and computing in general. If we provide a new paradigm of computing, we probably need new training and education material. 
These training materials can then be deployed on interactive and urgent HPC platforms, so that you get to practice what you are learning. 
User education and training is rich with opportunities to try, study, and observe how various teaching techniques, learning paths, and delivery methods can help users get proficient and successful as they use the HPC systems including interactive and urgent computing capabilities. 

For instance, typically HPC users are trained to use a given HPC system with either online documentation and/or classroom style tutorials in the classroom or over an online meeting. But some HPC centers have been developing and deploying asynchronous online courses (MOOCs) for such training. Which of these is most effective for students versus professionals who are learning to use the system? 
Further, if users start using the system with little or no experience, perhaps they should have access to limited resources. But they should be allocated enough resources to experience reasonably quick success on their project. Is this an effective path for getting them to reasonably quick proficiency to make their first HPC project successful? And does using the interactive and urgent features of the HPC system improve their likelihood of training success or hinder it? 

Looking even further ahead, can automated user support (like bots or HPC-Clippy!) help inexperienced users get up to speed quickly? How effective can automated user support suggest the next steps for getting an interactive or urgent job running, and how complex of a situation can such automated user support handle? And do the users even like having such automated support to help them use the system? These are just some of the research questions for education and training for interactive and urgent HPC. 

\subsection{Building Community}

Finally and above all, interactive and urgent HPC can only realize its full potential with the support of a collaborative community consisting of users, software developers, HPC center and vendor staff, and decision-makers with a stake in these capabilities.
Over the past several years we have taken our first steps building this community, leveraging birds-of-a-feather sessions and workshops at the major supercomputing conferences to come together, share, and document new tools, use cases, case studies, best practices, and lessons learned.
To build this community in order to address all of the above challenges, we need to continue these efforts but also identify new opportunities for funded collaboration, building new research partnerships, and interacting with each other more regularly and directly.
The authors will continue to organize interactive and urgent HPC workshops at the major supercomputing conferences and publishing associated proceedings to provide a voice for the community.
\section{Conclusion}
\label{sec:conclusion}

The case for increased consideration and development of interactive and urgent capabilities in HPC is compelling.
Our position is rooted in first-hand experience, observation, and scholarship at the interface between HPC and high-impact, time-sensitive workflows in science and engineering.
In this paper we have outlined the challenges and opportunities in advancing interactive and urgent capabilities, including policy, technology, scheduling, education, training, and community-building, and a number of priorities are emerging. 
It is in tackling, studying, learning, and sharing about advances on these challenges and opportunities that our community will go forth to have greater impact on scientific and technical research projects that require interactive and urgent HPC.

\section*{Acknowledgments}
% \rem{Add credit lines}
The authors thank Johannes Biermann (GDWG), Jens Henrik Goebbert (FZ-Juelich),  Monika Harlacher (SVA), Michael Ringenburg (Microsoft), and Rollin Thomas (LBL) for their valuable input to this paper. The authors also express their gratitude to all of the participants of the ISC-2022 Workshop and SC-2023 Birds-of-a-Feather sessions for sharing their experiences and insights in developing and deploying interactive and urgent HPC capabilities. 

% MIT LL co-authors are required to include this statement either in the Acknowledgements or as a foot note on page 1. 
DISTRIBUTION STATEMENT A. Approved for public release. Distribution is unlimited. This material is based upon work supported by the Department of the Air Force under Air Force Contract No. FA8702-15-D-0001. Any opinions, findings, conclusions or recommendations expressed in this material are those of the author(s) and do not necessarily reflect the views of the Department of the Air Force.

% Berkeley Lab co-authors need only reference Contract 31; no "use of NERSC resources here."
This work was supported by the Director, Office of Science, Office of Advanced Scientific Computing Research, of the U.S. Department of Energy under Contract No.~DE-AC02-05CH11231.

This work was supported by the Bundesministerium für Bildung und Forschung and the Niedersächsisches Ministerium für Wissenschaft und Kultur (KISSKI, Förderkennzeichen: 01IS22093A, and Nationales Hochleistungsrechnen an Hochschulen, NHR). 

\section*{Biographies}
\textbf{Albert Reuther} is a Senior Technical Staff member in the MIT Lincoln Laboratory Supercomputing Center in Lexington, MA 02421 USA. His research interests include interactive HPC, application specific computer architectures, and software/hardware co-design. Reuther received his PhD in Electrical and Computer Engineering from Purdue University. He is a member of the IEEE. Contact him at reuther@LL.mit.edu. 

% draft
\textbf{Nick Brown} is a Senior Research Fellow in the EPCC at the University of Edinburgh, Edinburgh, EH8 9BT, Scotland. His research interests include novel hardware, algorithmic techniques, and programming languages for high performance. Brown received his PhD from the University of Edinburgh. Contact him at n.brown@epcc.ed.ac.uk. 

\textbf{William Arndt} is a Computer Systems Engineer at Lawrence Berkeley National Laboratory in Berkeley, California, 94720, USA. His research interests include Bioinformatics algorithms, HPC system policy, and Transparent Checkpoint-Restart tools. Arndt received his PhD in Computer Science from the University of South Carolina. He is a member of IEEE. Contact him at warndt@lbl.gov.

\textbf{Johannes Blaschke} is a Computer Systems Engineer at The National Energy Research Scientific Computing Center (NERSC) at Lawrence Berkeley National Laboratory, California, 94720, USA. His research interests include using high-performance computing (HPC) for real-time data analysis in experimental and observational science, and future HPC workflows. Blaschke received his PhD in Physics from the Georg-August Universitaet Goettingen. Contact him at jpblaschke@lbl.gov. 

\textbf{Christian Boehme} is Deputy Head Of Computing Working Group at GWDG at Göttingen, 37073, Germany. His research interests include scheduling policies, HPC-as-a-Service, and scalable AI. Boehme received his Ph.D. in computational chemistry from the University of Marburg, Germany. 
Contact him at christian.boehme@gwdg.de. 

\textbf{Antony Chazapis} is a postdoctoral researcher at the Institute of Computer Science, FORTH at Heraklion, Crete, Greece. His research interests include Cloud-native software architectures, Cloud-HPC convergence, acceleration technologies, and distributed software-defined storage. He received his Ph.D. from the National Technical University of Athens, Greece. Contact him at chazapis@ics.forth.gr.

% draft
\textbf{Bjoern Enders} is a Data Science Workflows Architect at NERSC/LBL in Berkeley, CA, 94720, USA. His research interests focus on embedding HPC into data and compute workflows from experimental and observational facilities. He received his Ph.D. in physics from the Technical University of Munich. Contact him at benders@lbl.gov. 

\textbf{Robert Henschel} is Director of Research Engagement at Indiana University in Bloomington, Indiana, 47401, USA. His research interests include creating user interfaces and abstractions that simplify access to HPC systems for inexperienced users, as well as performance analysis of parallel scientific applications and workflows. Henschel received his master’s degree in computer science from Technische Universität Dresden, Germany. Contact him at henschel@iu.edu.

\textbf{Julian M. Kunkel} is Professor at the Universität Göttingen at Göttingen, 37027, Germany and a Deputy Head of the GWDG at Göttingen. His research interests include storage, performance analysis and management of HPC. Prof. Kunkel received his degree in computer science from the University of Hamburg. Contact him at julian.kunkel@gwdg.de.

\textbf{Maxime Martinasso} is the head of engineering at the Swiss National Supercomputing Centre at Lugano, 6900, Switzerland. His research interests include convergence of HPC and cloud technology, innovative access to HPC resources and performance modelling. Martinasso received his Ph.D. in computer science from the University of Grenoble, France. Contact him at maxime.martinasso@cscs.ch.

%Bibliography
\bibliographystyle{unsrt}  
\bibliography{references}

\begin{thebibliography}{10}

\bibitem{Zhou2020}
Xiao-Yun Zhou, Yao Guo, Mali Shen, and Guang-Zhong Yang.
\newblock Application of artificial intelligence in surgery.
\newblock {\em Frontiers of Medicine}, 14(4):417--430, 2020.

\bibitem{reuther2006high}
Albert Reuther and Suzy Tichenor.
\newblock High performance computing and competitiveness: Making the business
  case.
\newblock {\em Cyberinfrastructure Technology Watch Quarterly}, 2, 2006.

\bibitem{byun2020best}
Chansup Byun, Jeremy Kepner, William Arcand, David Bestor, Bill Bergeron, Vijay
  Gadepally, Michael Houle, Matthew Hubbell, Michael Jones, Andrew Kirby, Anna
  Klein, Peter Michaleas, Lauren Milechin, Julie Mullen, Andrew Prout, Antonio
  Rosa, Siddharth Samsi, Charles Yee, and Albert Reuther.
\newblock Best of both worlds: High performance interactive and batch
  launching.
\newblock pages 1--7. IEEE, 9 2020.

\bibitem{reuther2018scalable}
Albert Reuther, Chansup Byun, William Arcand, David Bestor, Bill Bergeron,
  Matthew Hubbell, Michael Jones, Peter Michaleas, Andrew Prout, Antonio Rosa,
  and Jeremy Kepner.
\newblock Scalable system scheduling for hpc and big data.
\newblock {\em Journal of Parallel and Distributed Computing}, 111, 2018.

\bibitem{turilli2018comprehensive}
Matteo Turilli, Mark Santcroos, and Shantenu Jha.
\newblock A comprehensive perspective on pilot-job systems.
\newblock {\em ACM Computing Surveys (CSUR)}, 51(2):1--32, 2018.

\bibitem{reuther2007technical}
Albert Reuther, Jeremy Kepner, Andy MCcabe, Julie Mullen, Nadya~T. Bliss, and
  Hahn Kim.
\newblock Technical challenges of supporting interactive hpc.
\newblock pages 403--409, 2007.

\bibitem{OpenOnDemand}
Dave Hudak, Doug Johnson, Alan Chalker, Jeremy Nicklas, Eric Franz, Trey
  Dockendorf, and Brian~L. McMichael.
\newblock Open ondemand: A web-based client portal for hpc centers.
\newblock {\em Journal of Open Source Software}, 3(25):622, 2018.

\bibitem{MaterialClouds}
Leopold Talirz, Snehal Kumbhar, Elsa Passaro, Aliaksandr~V. Yakutovich, Valeria
  Granata, Fernando Gargiulo, Marco Borelli, Martin Uhrin, Sebastiaan~P. Huber,
  Spyros Zoupanos, Carl~S. Adorf, Casper~Welzel Andersen, Ole Schütt, Carlo~A.
  Pignedoli, Daniele Passerone, Joost VandeVondele, Thomas~C. Schulthess,
  Berend Smit, Giovanni Pizzi, and Nicola Marzari.
\newblock Materials cloud, a platform for open computational science.
\newblock {\em Scientific Data}, 7(1), sep 2020.

\bibitem{vcluster}
Sadaf~R Alam, Miguel Gila, Mark Klein, Maxime Martinasso, and Thomas~C
  Schulthess.
\newblock Versatile software-defined hpc and cloud clusters on alps
  supercomputer for diverse workflows.
\newblock {\em The International Journal of High Performance Computing
  Applications}, 37(3-4):288--305, 2023.

\bibitem{blaschke2021realtime}
Johannes~P. Blaschke, Aaron~S. Brewster, Daniel~W. Paley, Derek Mendez, Asmit
  Bhowmick, Nicholas~K. Sauter, Wilko Kröger, Murali Shankar, Bjoern Enders,
  and Deborah Bard.
\newblock Real-time xfel data analysis at slac and nersc: a trial run of
  nascent exascale experimental data analysis, 2021.

\bibitem{mullen2018lessons}
Julia Mullen, Albert Reuther, William Arcand, Bill Bergeron, David Bestor,
  Chansup Byun, Vijay Gadepally, Michael Houle, Matthew Hubbell, Michael Jones,
  Anna Klein, Peter Michaleas, Lauren Milechin, Andrew Prout, Antonio Rosa,
  Siddharth Samsi, Charles Yee, and Jeremy Kepner.
\newblock {\em Lessons Learned from a Decade of Providing Interactive,
  On-Demand High Performance Computing to Scientists and Engineers}, pages
  655--668.
\newblock Springer, Cham, 6 2018.

\end{thebibliography}

\end{document}